\documentclass{article}
\usepackage{graphicx} % Required for inserting images

\usepackage{cite}
\usepackage{xcolor}
\usepackage[colorlinks=true,linkcolor=blue, citecolor=red, urlcolor=magenta, bookmarks]{hyperref}

\usepackage{authblk}

\usepackage{physics}
\usepackage{amsmath}
\usepackage{amsfonts}
\usepackage{dsfont}
\usepackage{tikz}
\usepackage{circuitikz}
\usetikzlibrary{arrows,decorations.pathreplacing}

\tikzset{meter/.append style={draw, inner sep=10, rectangle, font=\vphantom{A}, minimum width=30, line width=.8,
 path picture={\draw[black] ([shift={(.1,.3)}]path picture bounding box.south west) to[bend left=50] ([shift={(-.1,.3)}]path picture bounding box.south east);\draw[black,-latex] ([shift={(0,.1)}]path picture bounding box.south) -- ([shift={(.3,-.1)}]path picture bounding box.north);}}}

\newcommand{\hg}[2]{\,\mbox{}_{#1}F_{ #2}\!}

\newcommand{\argu}[3]{\left(\begin{array}{c} #1\\#2\end{array} ; #3\right)}

\usepackage[a4paper, total={6in, 9in}]{geometry}

\begin{document}

\title{A dynamical algebra of protocol-induced transformations on Dicke states}

\author[1]{Pierre-Antoine Bernard\footnote{ {\normalsize Corresponding author:
\texttt{bernardpierreantoine@outlook.com}}}}
\author[1,2]{Luc Vinet}

\affil[1]{\it Centre de Recherches Math\'ematiques (CRM), Universit\'e de Montr\'eal, P.O. Box 6128, Centre-ville
Station, Montr\'eal (Qu\'ebec), H3C 3J7, Canada,}

\affil[2]{\it IVADO,  Montréal (Québec), H2S 3H1, Canada.}

\date{December 23, 2024}

\maketitle
\vspace{-0.5cm}
\begin{abstract}
Quantum $n$-qubit states that are totally symmetric under the permutation of qubits are essential ingredients of important algorithms and applications in quantum information. Consequently, there is significant interest in developing methods to prepare and manipulate Dicke states, which form a basis for the subspace of fully symmetric states.
Two simple protocols for transforming Dicke states are considered. An algebraic characterization of the operations that these protocols induce is obtained in terms of the Weyl algebra $W(2)$ and $\mathfrak{su}(2)$. Fixed points under the application of the combination of both protocols are explicitly determined. Connections with the binary Hamming scheme, the Hadamard transform, and Krawtchouk polynomials are highlighted.
\end{abstract}

\noindent \textbf{Keywords:} Quantum protocols, Dicke states, dynamical and symmetry algebras, Hamming scheme, Krawtchouk polynomials. \\
\noindent \textbf{MSC classes}: {\tt 81P65}, {\tt 22E60}, {\tt 81V72}, {\tt 05E30}.

%\tableofcontents

\section{Introduction}
Dicke states $|D_n^{i} \rangle$ are the coherent superpositions of all the $n$-qubit state vectors with $i$ single qubit states $|1\rangle$ and $n-i$ single qubit states $|0\rangle$. They are totally symmetric under the permutations of the qubits. More precisely, let $V=\{0,1\}^n$ be the set of binary sequences $x=(x_1, x_2, \dots, x_n)$ of length $n$ and $\partial(x,y)$ be the Hamming distance between $x,y \in V$ defined by
\begin{equation}
    \partial(x,y) = |\{ i \in \{1, 2, \dots, N\} \ |\ x_i \neq y_i\}|.
\end{equation}
To each sequence $x$ is associated an orthonormalized vector $|x\rangle$ given by 
\begin{equation}
    \ket{x} = \ket{x_1}\otimes \dots \otimes \ket{x_n} \in \mathbb{C}^{2^N}, %\quad \text{with} \quad \ket{0} = \binom{1}{0} \quad \text{and} \quad \ket{1} = \binom{0}{1},
\end{equation}
where $\ket{0} = \binom{1}{0}$ and $\ket{1} = \binom{0}{1}$ in the computational basis. By definition,

\begin{equation}
    \ket{D_n^{i}} = \frac{1}{\sqrt{\binom{n}{i}}}\sum_{\substack{x \in V \\ \partial(x, \boldsymbol{0}) = i}} \ket{x}, \quad i \in \{0,1,...,n\}.\label{Dicke}
\end{equation}
Dicke states
%\footnote{See \cite{raveh2024dicke} for their matrix product state (MPS) representation.}
were first introduced to study coherent radiation arising from the correlated motion of gas particles \cite{dicke1954coherence}, and have since found applications in various contexts, particularly in quantum information science. They play a central role in numerous quantum algorithms, including the quantum approximate optimization algorithm (QAOA) \cite{farhi2014quantum}. As a result, significant efforts continue to be devoted to the efficient construction of these states \cite{childs2000finding,bartschi2019deterministic,buhrman2024state,piroli2024approximating,nepomechie2023qudit,yu2024efficient,liu2024low}.

%These Dicke states\footnote{See \cite{raveh2024dicke} for their MPS presentation} are of importance in various contexts and in particular they feature prominently in many quantum algorithms such as the quantum approximate optimization one \cite{farhi2014quantum}. Hence, understandably, significant efforts continue to be devoted to the efficient construction of these states \cite{childs2000finding,bartschi2019deterministic,buhrman2024state,piroli2024approximating,nepomechie2023qudit,yu2024efficient,liu2024low}. 

In the context of these studies, it is of interest to examine the design of gates and operations that transform Dicke states among themselves with the effect of adding or subtracting qubits, that is of changing $n$. This has been considered in \cite{kobayashi2014universal} allowing for the Dicke state to be shared by two parties. The authors provided universal gates that can be independently implemented by one of the two parties, making possible the transformation of a given Dicke state into a target one. The present paper will bear on the task of manipulating Dicke states in the simplest possible way. First, it will be assumed that the resource is not divided. Second, it will discuss protocols for two basic operations where only one qubit is either added or subtracted. With these transformations viewed as building blocks, our goal will be to characterize the set of operations they enable by bringing to light the algebraic structure that they entail. The Weyl algebra $W(2)$ and its $\mathfrak{su}(2)$ will be seen to appear. %\color{red}It will also be observed that the composition of these two operations yields an action on Dicke states that corresponds to the dynamics on a path that exhibits perfect state transfer.\color{black}

Generators that transform among themselves the degenerate states of a certain Hamiltonian are said to span the \textit{symmetry algebra} of the system.  A \textit{dynamical algebra} (typically non-compact) should include the symmetry algebra and be such that it is possible to insert all degenerate subspaces in one of its irreducible representation. The Dicke states on $n$ qubits are known to form a basis for an irreducible $(n+1)$-dimensional representation of $\mathfrak{su}(2)$ and can be seen as spanning the degenerate subspace associated with the eigenvalue $n$ of a Hamiltonian counting the number of qubits supporting totally symmetric states. With this nomenclature in mind, the direct sum of two copies $W(2) = W(1) \oplus W(1)$ of the Weyl algebra will be seen to emerge as the dynamical algebra generated by the transformations on Dicke states induced by the two protocols. The combined transformations produced by the two protocols, which preserve the total number of qubits, are identified within this framework as elements of the symmetry algebra, forming a representation of (the complexification of) $\mathfrak{su}(2)$.
%The Dicke states on $n$ qubits are known to form a basis for an irreducible $(n+1)$-dimensional representation of $\mathfrak{su}(2)$ (we shall in fact consider in the following its complexification $ \mathfrak{sl}_2(\mathbb{C})$) which thus preserves $n$ and can be viewed as a symmetry algebra.

%the semi-direct product of the one-dimensional Weyl algebra with $\mathfrak{sl}_2 (\mathbb{C})$ will present itself as the dynamical algebra realized by our set of basic operations on Dicke states.

This report will unfold as follows. We shall set the stage in the next section with a description of the two simple protocols that involve measurements and one-qubit gates. Before moving to the description of their effects on symmetric states, we shall review in Section \ref{sec:3} the connection between Dicke states and $\mathfrak{su}(2)$ in the framework of the hypercube and the Hamming association scheme \cite{bernard2024q}. The ubiquity of the Krawtchouk polynomials will be manifest and their role in the Hadamard transform of the Dicke states will be stressed. Section \ref{sec:4} will detail the outcomes of Protocol 1 and Protocol 2 when they are successful, highlighting that the respective transformations remove or add one qubit to Dicke states. Additionally, the transformation resulting from the combination of the two protocols will be described. Section \ref{sec:5} will discuss the algebraic structure that these operations realize. It will be observed that Protocols 1 and 2 generate a representation of the Weyl algebra $W(2)$, and that their combination gives a representation of the complexification of $\mathfrak{su}(2)$. 
Section \ref{sec:6} will explore potential applications of this framework, including the preparation of Dicke states. It will also examine the asymptotic states arising from iterative applications of the two protocols. A conclusion, \ref{sec:7}, and two appendices, one, \ref{app:A}, on Krawtchouk polynomials and the other, \ref{App:B}, on the measurement of total angular momentum will complete the paper.

\section{General framework: the protocols}\label{sec:2}
We shall focus on quantum states $\ket{\psi}$ in $n$ qubits that are totally invariant under permutations of the qubits and that have hence the following decomposition in terms of Dicke states:
\begin{equation}\label{def:nqb}
    \ket{\psi} = \sum_{i = 0}^n \psi_{n,i} \ket{D_n^i} , \quad  \sum_{i =0}^n |\psi_{n,i}|^2 = 1.
\end{equation}
%with $ \ket{D_n^i}$ the so-called $i$-th Dicke state on $n$ qubits,
%\begin{equation}
   % \ket{D_n^i} = \binom{n}{i}^{-1/2} \sum_{\substack{\boldsymbol{x} \in \mathbb{Z}_2^n \\ |\boldsymbol{x}| = i}}  \ket{\boldsymbol{x}}.
%\end{equation}
Consider the Hilbert space $(\mathbb{C}^2)^{\otimes n}$ of $n$ qubits. We denote by $\mathcal{D}_n$ its subspace spanned by Dicke states, and by $\mathcal{D}$ their direct sum:
\begin{equation}
    \mathcal{D} = \bigoplus_{n = 0}^{\infty} \mathcal{D}_n, \quad \mathcal{D}_n  = \text{span}\{ \ket{D_n^i}\, |\, i = 0,1,\dots, n\}\subset (\mathbb{C}^2)^{\otimes n}.
\end{equation}
We are interested in the transformations of the quantum states $\ket{\psi}$ under two simple protocols based on measurements and single-qubit gates:
\begin{itemize}
    \item \textit{Protocol 1 (Measuring a qubit):} An arbitary qubit from the $n$ qubits encoding the state $\ket{\psi}$ is selected. A single qubit gate is applied on this qubit, and a measurement is realized. If the result is $\ket{0}$, the protocol is a success and the output is given by the remaining $n-1$ qubits state. Otherwise, the protocol fails and the state of the whole system is discarded.  

    \item \textit{Protocol 2 (Measuring total angular momentum with an additional qubit)}: An additional qubit in the state $\ket{0}$ is introduced to the system of $n$ qubits, initially in the state $\ket{\psi}$. A single-qubit unitary gate is then applied to this new qubit, followed by a measurement of the total angular momentum $j$ of the combined $n+1$ qubits system (see Appendix \ref{App:B} for a circuit implementation of this measurement using SWAP gates). If the measured angular momentum is maximal, i.e. $j = (n+1)/2$, the protocol succeeds, and the resulting $n+1$ qubit state is retained. Otherwise, the protocol fails, and the state is discarded.
\end{itemize}

In the following sections, we analyze the dynamical algebra associated with the transformations induced by these protocols when they succeed. %To this end, we first revisit the connection between Dicke states, the hypercube graph, and the binary Hamming scheme.

\section{Dicke states and the Hamming association scheme}\label{sec:3}

The Dicke states $|D_n^{i}\rangle$ have a natural connection with $\mathfrak{su}(2)$, as well as with the hypercube $Q_n$ which is a graph of the (binary) Hamming association scheme. This is useful to have in mind and will be reviewed in this section.

The set of $n$-bit strings $V$ corresponds to the vertices of $Q_n$. There is an edge between vertices at Hamming distance $1$. This is represented by the adjacency matrix $A$ with elements:

\begin{equation}\label{def:A}
    \bra{x}A\ket{y} = \left\{
	\begin{array}{ll}
		1  & \mbox{if } \partial(x,y) =1  \\
		0 & \mbox{otherwise. }
	\end{array}
\right.
\end{equation}
The distance matrices $A_i$, with $i= 0,\dots,n$, are similarly defined with their non-zero entries being $\langle x|A_i|y\rangle = 1$ if $\partial (x,y) = i$. We see that
\begin{align}
    &A_0 = I, \label{A1}\\
    &A=A^{\top},\label{A2}\\
    &A_0+A_1+\dots+A_n=J \label{A3}
\end{align}
with $J$ the all ones matrix. Note that $A_1=A$. These matrices obey in addition the Bose-Mesner algebra relation
\begin{equation}
    A_iA_j=\sum_{k=0}^n p_{ij}^k A_k \label{A4},
\end{equation}
where the intersection parameters $p_{ij}^k$ count the number of $z$ such that $\partial (x,z)=i$ and $\partial(y,z)=j$ if $\partial (x,y)=k$.

In general a symmetric association scheme is defined as a family of  $(0,1)$-matrices $\{A_0, A_1, \dots, A_n\}$ verifying conditions \eqref{A1}, \eqref{A2}, \eqref{A3} and \eqref{A4}. The binary Hamming scheme is the example arising as described above from the hypercube. This special graph is further known to be distance-regular. This implies that $p_{i1}^k$ is a tridiagonal matrix. Precisely, in the case of the hypercube, we have
\begin{equation}
    AA_i=(i+1)A_{i+1} + (n-i+1)A_{i-1}.\label{3t}
\end{equation}
The combinatorial determination of the parameters $p_{i1}^{i\pm1}$ and $p_{i1}^{i}$ is straightforward and is explained in \cite{bernard2018graph} for instance. In view of this three-term recurrence relation, it follows that $A_i$ is a polynomial of degree $i$ in $A$. Whenever such a relation between $A_i$ and $A$ happens, the association scheme is said to be P-polynomial. Specifically, for the Hamming scheme, one finds from equation \eqref{3t} that
\begin{equation}
    A_i= \binom{n}{i} K_i\left(\frac{n-A}{2};\frac{1}{2},n\right),
\end{equation}
where $K_i$ are the Krawtchouk polynomials defined in the Appendix \ref{app:A}. It follows from the definition of Dicke states in equation \eqref{Dicke} and the definition of the distance matrices $A_i$ that they are related by the following expression:
\begin{equation}\label{D&Ai}
    |D_n^{i}\rangle = \frac{1}{\sqrt{\binom{n}{i}}} A_i|\textbf{0}\rangle  = {\sqrt{\binom{n}{i}}} K_i\left(\frac{n-A}{2};\frac{1}{2},n\right)|\textbf{0}\rangle.   
\end{equation}

\subsection{Connection with $\mathfrak{su}(2)$}

It is also relevant to introduce the dual adjacency matrix $A^*$ which for the binary Hamming scheme is defined by
\begin{equation}
    A^*|x\rangle=  \left(n-2\partial(x,\bold{0})\right)|x\rangle,
\end{equation}
where $\bold{0} = (0,\dots,0)$. The ajacency and dual adjacency matrices $A$ and $A^*$ yield a natural connection between the hypercube and the Lie algebra $\mathfrak{su}(2)$, with generators $j^x$, $j^y$ and $j^z$ obeying $[j^x,j^y]=ij^z$ and cycl. Obviously the vectors $\ket{0}$ and $\ket{1}$ form a basis for the fundamental spin-$\frac{1}{2}$ representation of this algebra in which the generators are represented as 
\begin{equation}
    j^a \rightarrow \frac{1}{2} \sigma^a, \quad a = x,y,z, 
\end{equation}
with $\sigma^x$, $\sigma^y$ and $\sigma^z$ the Pauli matrices
\begin{equation}
    \sigma^x=\begin{pmatrix}
        0&1\\
        1&0
    \end{pmatrix}, 
    \quad \sigma^y=\begin{pmatrix}
        0&-i\\
        i&0
    \end{pmatrix},
    \quad \sigma^z=\begin{pmatrix}
        1&0\\
        0&-1
    \end{pmatrix}.
\end{equation}
Higher spin representations are constructed from the fundamental representation through repeated applications of the coproduct, $\Delta: \mathfrak{su}(2)\rightarrow \mathfrak{su}(2)\otimes \mathfrak{su}(2)$, defined by
\begin{equation}
    \Delta(j^{a}) = j^{a}\otimes I + I \otimes j^{a}.
\end{equation}
In particular, $n-1$ applications of the coproduct $\Delta$ on the fundamental representation yield a (reducible) representation of $\mathfrak{su}(2)$, which acts on the Hilbert space $(\mathbb{C}^2)^{\otimes n}$ associated with $n$ spin-$\frac{1}{2}$ particles. In this case the abstract basis elements of $\mathfrak{su}(2)$ are represented by $2^n \times 2^n $ matrices $J^{a}$ given by
\begin{equation}\label{eq:defJ}
    j^a \rightarrow J^{a}= \Delta^{(N-1)} \left(\sigma^a/2\right)  = \sum_{i=1}^n \underbrace{I\otimes...\otimes I}_{i-1\text{ times}}\otimes\ \frac{\sigma^{a}}{2}\otimes \underbrace{I\otimes...\otimes I}_{N-i\text{ times}},
\end{equation}
with $a = x,y,z$. A connection between $\mathfrak{su}(2)$ and the hypercube is established by noting that $A$ and $A^*$ can be expressed respectively in terms of the representations of $j^x$ and $j^z$ in $(\frac{1}{2})^{\otimes n}$ as per \eqref{eq:defJ}; indeed one has:
\begin{equation} \label{AA*}
    A=2J^x \quad \text{and} \quad A^*=2J^z.
\end{equation}
%from equation \eqref{eq:defj},
%\begin{equation}
   % A = 2 j^x = \sum_{i=1}^N \underbrace{I\otimes...\otimes I}_{i-1\text{ times}}\otimes\ \sigma_x\otimes \underbrace{I\otimes...\otimes I}_{N-i\text{ times}}, \label{Adj}
%\end{equation}

%\begin{equation}
    %A^* = 2 j^z = \sum_{i=1}^N \underbrace{I\otimes...\otimes I}_{i-1\text{ times}}\otimes\ \sigma_z\otimes \underbrace{I\otimes...\otimes I}_{N-i\text{times}}.\label{Adjd}
%\end{equation}
We thus see that the adjacency matrix $A$ and its dual $A^*$ belong to the representation of $\mathfrak{su}(2)$ given by the $n$-fold tensor product of the spin-$\frac{1}{2}$ representation. For the Hamming scheme, $\mathfrak{su}(2)$ happens to be the Terwilliger algebra that can be attached to association schemes \cite{go2002terwilliger, bernard2023entanglement}.

Recalling that $\sigma^-|0\rangle=|1\rangle$ with $\sigma^{\pm}=\frac{1}{2}(\sigma^x\pm i\sigma^y)$, it is readily found that the Dicke states \eqref{Dicke} are given by
\begin{equation}
    |D_n^{i}\rangle = \frac{1}{i!\sqrt{\binom{n}{i}}}\left(\Delta^{(n-1)}(\sigma^-)\right)^i |\textbf{0} \rangle, \quad i \in \{0, 1, \dots, n\} \label{Dickedelta} .
\end{equation} 
Using the expressions arising from \eqref{eq:defJ} for the adjacency matrix $A=2J^x$ and its dual $A^*=2J^z$, it is straightforward to compute their actions on the Dicke states  given by \eqref{Dickedelta} to find 
\begin{equation}
    A |D_n^{i}\rangle  = \sqrt{(i+1)(n-i)} |D_n^{i+1}\rangle + \sqrt{i(n-i+1)} |D_n^{i-1}\rangle, \label{AonD}
\end{equation}
\begin{equation}
    A^*|D_n^{i}\rangle = (n-2i) |D_n^{i}\rangle\label{A*onD},
\end{equation}
with $i=0,\dots,n.$ This indicates that the Dicke states $|D_n^{i}\rangle$ transform irreducibly under a spin-$\frac{n}{2}$ representation of $\mathfrak{su}(2)$. The identification with the angular momentum states $|j,m\rangle$ has $j=\frac{n}{2}, m=\frac{n}{2} - i$; note that the eigenvalue of $A^*$ on those states is $2m$ under this correspondence. Relations \eqref{AonD} and \eqref{A*onD} also show that restricting $\mathbb{C}^{2^n}$ to the span of the Dicke states $|D_n^{i}\rangle$,  with $i=0\dots,n$,
picks the highest spin representation $(j=\frac{n}{2})$ in the irreducible decomposition of the $n$-fold tensor product $(\frac{1}{2})^{\otimes n}$. In terms of graphs, it corresponds to  considering a quotient graph of the hypercube, down to the weighted path that admits perfect state transfer \cite{christandl2005perfect, bernard2018graph}.

\subsection{Hadamard transform}

Here, we recall the definition of the Hadamard transform and discuss its connection to the adjacency and dual adjacency matrices of the hypercube $Q_n$. Let $f$ be a function from $V$ to $\mathbb{C}$. Consider the Hadamard-Walsh gate $H^{\otimes n}$ with 
\begin{equation}\label{H}
    H=\frac{1}{\sqrt{2}}\begin{pmatrix}
        1&1\\
        1&-1
    \end{pmatrix}
    =\frac{\sigma^x + \sigma^z}{\sqrt{2}} = (-i) e^{\frac{i \pi}{2} \frac{\sigma^x + \sigma^z}{\sqrt{2}}}.
\end{equation}
The action of $H^{\otimes n}$ on a state in $(\mathbb{C}^2)^{\otimes n}$ is
\begin{equation}
    H^{\otimes n}\sum_{x \in V} f(x)|x\rangle = \sum_{x \in V} \tilde{f}(x) |x\rangle,
\end{equation}
where $\tilde{f}(x)$ is the Hadamard transform of $f(x)$:
\begin{equation}
    \tilde{f}(x)=\frac{1}{\sqrt{2^n}}\sum_{y \in V} (-1)^{x.y}f(y).
\end{equation}
In view of \eqref{eq:defJ}, \eqref{AA*} and \eqref{H}, one has that $H^{\otimes n} =(-i)^n e^{\frac{i \pi}{2} \frac{A + A^*}{\sqrt{2}}}$.  It is then clear from the $\mathfrak{su}(2)$ context that
\begin{equation}
    H^{\otimes n}AH^{\otimes n}=A^*.
\end{equation}
In other words we see that $H^{\otimes n}$ diagonalizes the matrix $A$ that has the same spectrum $\{n-2i, \, |\, i=0,\dots,n\}$ as $A^*$. Consider this diagonalization on the subspace of totally symmetric states and let
\begin{equation}
    A|F_n^{i}\rangle = (n-2i)|F_n^{i}\rangle,
    \label{eig}
\end{equation}
with
\begin{equation}
    |F_n^{i}\rangle = \sum_{k=0}^n \langle D_n^{k}|F_n^{i}\rangle |D_n^{k}\rangle.
\end{equation}
From $\langle D_n^{k}|A|F_n^{i}\rangle = (\langle D_n^{k}|A)|F_n^{i}\rangle $, equations \eqref{eig} and \eqref{AonD} for the action of $A$, and the three term recurrence relation of the Krawtchouk polynomials one finds that
\begin{equation}
    \langle D_n^{k}|F_n^{i}\rangle = \sqrt{\binom{n}{i}\binom{n}{k}} K_k\left(i;\frac{1}{2},n\right) = \sqrt{\binom{n}{i}\binom{n}{k}} K_i\left(k;\frac{1}{2},n\right).
\end{equation}
where the duality ($K_i(k; p,n) = K_k(i;p,n)$) of the Krawtchouk polynomials has been used in the last equality.

%Either by a combinatorial argument or by observing that the action of $A$ on both vectors is the same given \eqref{3t}, one has that 
%\begin{equation}
%    |D_n^{i}\rangle = \frac{1}{\sqrt{\binom{n}{i}}} A_i|\textbf{0}\rangle .   
%\end{equation}
We may also directly compute the action of the Hadamard-Walsh gate $ H^{\otimes n}$ on Dicke states which offers an alternative way of computing the eigenstates of $A$. The starting point is \eqref{D&Ai} that relates the Dicke states $|D_n^{i}\rangle$ to the application of $A_i$ on $|0\rangle^{\otimes n}$. Note that \eqref{AonD} allows to double check \eqref{D&Ai} by observing that the action of $A$ on both sides of \eqref{D&Ai} is the same in view of \eqref{3t}.
So one has
\begin{equation}
    \begin{split}
     H^{\otimes n}|D_n^{i}\rangle =& \frac{1}{\sqrt{\binom{n}{i}}} H^{\otimes n} A_i|\textbf{0}\rangle\\
     =&\sqrt{\binom{n}{i}}H^{\otimes n}K_i\left(\frac{n-A}{2};\frac{1}{2},n\right)|0\rangle\\
     =&\sqrt{\binom{n}{i}}K_i\left(\frac{n-A^*}{2};\frac{1}{2},n\right)H^{\otimes n}|0\rangle\\
     =&\sum_{x \in V} \sqrt{\binom{n}{i}}K_i\left(\frac{n-A^*}{2};\frac{1}{2},n\right)|x\rangle\\
     =&\sum_{k=0}^n \sqrt{\binom{n}{i}\binom{n}{k}}K_i\left(k;\frac{1}{2},n\right) |D_n^{k}\rangle.
    \end{split}
\end{equation}
which indeed matches with the results from the straigthforward diagonalization described above. We therefore have that the Hadamard transform of the symmetric states \eqref{def:nqb} reads:
\begin{equation}
    H^{\otimes n}|\psi\rangle= \sum_{i=0}^n \tilde\psi_{n,i}|D_n^{i}\rangle \quad \text{with} \quad \tilde\psi_{n,i} = \sum_{k=0}^n \sqrt{\binom{n}{i}\binom{n}{k}}K_i\left(k;\frac{1}{2},n\right)\psi_{n,k}.
    \end{equation}
Some of the above considerations relate to the construction of superpositions of Dicke states within a certain distance from a reference bit string \cite{farhi2024efficiently}.

\section{Transformations of symmetric states}\label{sec:4}

In this section, we show that Protocols 1 and 2 map symmetric states to symmetric states when they succeed, thereby preserving the space $\mathcal{D}$ spanned by all Dicke states. We also provide an explicit description of the transformations induced by these maps and their composition.

\subsection{The transformation induced by Protocol 1}

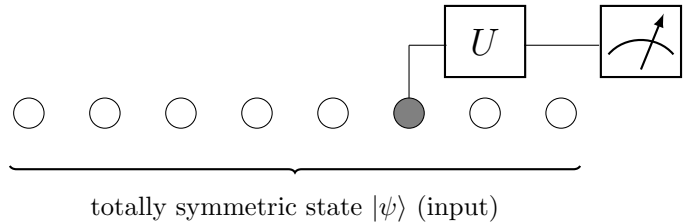
\begin{figure}[h] \centering
\begin{tikzpicture}
\draw  (0,0) circle (0.2cm);
\draw  (1,0) circle (0.2cm);
\draw  (2,0) circle (0.2cm);
\draw  (3,0) circle (0.2cm);
\draw  (4,0) circle (0.2cm);

\draw (5,0) -- (5,0.9);
\draw (5,0.9) -- (7.5,0.9);

\draw[fill=gray]  (5,0) circle (0.2cm);
\draw  (6,0) circle (0.2cm);
\draw  (7,0) circle (0.2cm);

\node[meter] (meter) at (8.05,0.95) {};

\node[meter] (meter) at (6,0.95) {};

\draw[fill=white, color = white] (5.5,1.4) rectangle (6.5,0.5);

\node[] at (6,0.95) {\begin{Large}$U$\end{Large}};

\draw [
    thick,
    decoration={
        brace,
        mirror,
        raise=0.5cm
    },
    decorate
] (-.25,-0.2) -- (7.25,-0.2) 
node
[pos=0.5,anchor=north,yshift=-0.75cm] {totally symmetric state $\ket{\psi}$ (input)}; 
\end{tikzpicture}
\caption{Schematic representation of Protocol $1$. The gray and white circles represent the qubits supporting the initial totally symmetric state $\ket{\psi}$. One qubit (gray) is selected arbitrarily. If the measurement following the application of the gate $U = U(\alpha, \beta)$ yields zero, the protocol succeeds, outputting the state of the remaining (white) qubits. } 
\end{figure}

Let $\ket{\psi}$ be a symmetric $n$-qubit state as given in \eqref{def:nqb}. Protocol 1 involves applying a unitary single-qubit gate to a selected target qubit and performing a measurement on it. Since we are considering states that are invariant under permutations, we can, without loss of generality, assume that this protocol is applied to the last qubit. Thus, it is useful to consider the following decomposition of the Dicke state $\ket{D_n^i}$, derived straighforwardly from a combinatorial argument
\begin{equation}
     \ket{D_n^i} = \sqrt{\frac{i}{n}} \ket{D_{n-1}^{i-1}} \otimes \ket{1} + \sqrt{\frac{n-i}{n}} \ket{D_{n-1}^{i}}\otimes \ket{0}.
\end{equation}
An arbitrary single gate qubit can be parametrized (up to a global phase factor) in terms of complex numbers $\alpha$ and $\beta$ as
\begin{equation}
    U(\alpha,\beta) = \begin{pmatrix}
        \alpha & \beta \\
        \beta^* & -\alpha^*
    \end{pmatrix}, \quad |\alpha|^2 + |\beta|^2 = 1 \quad \alpha, \beta \in \mathbb{C}.
\end{equation}
Under the application of this gate to the last qubit of a system of $n$ qubits, the Dicke state $\ket{D_n^i}$ is mapped to the following state
\begin{equation}\label{eq:ugate1}
\begin{split}
      (\mathds{1}^{\otimes (n-1)} \otimes U(\alpha,\beta) )\ket{D_n^i} & = \left(\alpha  \sqrt{\frac{n-i}{n}}\ket{D_{n-1}^{i}}+ \beta \sqrt{\frac{i}{n}}\ket{D_{n-1}^{i-1}}  \right) \otimes  \ket{0} \\
      & + \left(\beta^*  \sqrt{\frac{n-i}{n}}\ket{D_{n-1}^{i}}- \alpha^* \sqrt{\frac{i}{n}}\ket{D_{n-1}^{i-1}}  \right) \otimes  \ket{1}. 
\end{split}
\end{equation}
According to the first protocol, the next step is to perform a measurement on the last qubit. The protocol is considered a success only if the measurement outcome is $0$. From equation \eqref{eq:ugate1}, which describes the action of the single-qubit gate on each Dicke state, it follows that a successful outcome of this protocol results in the following state on the remaining $n-1$ qubits:
\begin{equation}
    \ket{\psi} \longrightarrow \ket{\psi'} = \kappa \sum_{i = 0}^{n-1} \psi_{n-1,i}' \ket{D_{n-1}^{i}},
\end{equation}
with $\kappa$ a normalization constant and $\psi_{n-1,i}'$ the coefficients expressed as follows in terms of the coefficients $\psi_{n,i}$ defining the initial symmetric state $\ket{\psi}$
\begin{equation}\label{def:ac1}
     \psi_{n-1,i}' = \alpha \sqrt{n-i} \, \psi_{n,i} + \beta  \sqrt{i+1} \, \psi_{n,i+1}.
\end{equation}
The transformation induced by a successful application of Protocol 1 is thus given, up to a normalization constant $\kappa$, by the linear operator $P_1(\alpha, \beta): \mathcal{D} \to \mathcal{D}$, whose action on a given Dicke state is
\begin{equation}
    P_1(\alpha,\beta)  \ket{D_{n}^i} = \alpha \sqrt{n-i}\ket{D_{n-1}^i} + \beta\sqrt{i}\ket{D_{n-1}^{i-1}}.
\end{equation}

\subsection{The transformation induced by Protocol 2}

\begin{figure}[h] \centering
\begin{tikzpicture}
\draw  (0,0) circle (0.2cm);
\draw  (1,0) circle (0.2cm);
\draw  (2,0) circle (0.2cm);
\draw  (3,0) circle (0.2cm);
\draw  (4,0) circle (0.2cm);

\draw (7,0) -- (7,0.9);
\draw (7,0.9) -- (7.5,0.9);

\draw  (5,0) circle (0.2cm);
\draw  (6,0) circle (0.2cm);
\draw[fill=gray]  (7,0) circle (0.2cm);

%\node[meter] (meter) at (8.05,0.95) {};

\node[meter] (meter) at (8,0.95) {};

\draw[fill=white, color = white] (7.5,1.4) rectangle (8.5,0.5);

\node[] at (8,0.95) {\begin{Large}$U$\end{Large}};

\draw [
    thick,
    decoration={
        brace,
        raise=0.5cm
    },
    decorate
] (-.25,0) -- (6.25,0) 
node
[pos=0.5,anchor=south,yshift=0.75cm] {totally symmetric state $\ket{\psi}$ (input)}; 

\draw [
    thick,
    decoration={
        brace,
        mirror,
        raise=0.5cm
    },
    decorate
] (-.25,-0.2) -- (7.25,-0.2) ;

\node[meter] (meter) at (3.5,-1.5) {};
\node[align = center] at (6.25,-1.5) {total angular\\ momentum measurement};

\end{tikzpicture}
\caption{Schematic representation of Protocol $2$. The white circles represent the qubits supporting the initial totally symmetric state $\ket{\psi}$. A single additional qubit (gray) is introduced and acted upon with a gate $U = U(\gamma, \delta^*)$. A measurement of the total angular momentum of the $n + 1$ qubits (gray and white) is then performed. If the measurement yields the maximal value, the protocol succeeds, outputting the state of the $n + 1$ qubits. Otherwise, the protocol fails, and the state is discarded.
} 
\end{figure}
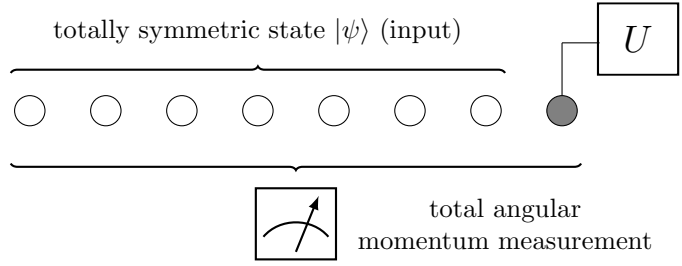

We now consider the second protocol where an additional qubit in a state $\ket{0}$ is added to the system and acted upon with a single qubit gate, i.e.
\begin{equation}
    \ket{\psi} \longrightarrow \ket{\psi}\otimes U(\gamma,\delta^*)\ket{0} = \ket{\psi}\otimes (\gamma \ket{0} + \delta \ket{1}).
\end{equation}
The overlap between this state and the symmetric Dicke states defined on $n+1$ qubits can be computed using standard angular momentum theory. Indeed, as mentioned in \ref{sec:3}, the Dicke state $\ket{D_n^i}$ can be identified as the common eigenvector $\ket{j_1, m_1}$, with $j_1 = n/2$ and $m_1 = n/2 - i$, of the Casimir operator $\boldsymbol{j}^2 = (j^x)^2 + (j^y)^2 + (j^z)^2 $ and the $z$-component $j^z$ of the total angular momentum operator on the corresponding irreducible submodule of $(\frac{1}{2})^{\otimes n}$. Similarly, the state $\ket{0}$ (or $\ket{1}$) can be identified with the eigenvectors $\ket{j_2, m_2}$, with $j_2 = 1/2$ and $m_2 = 1/2$ (or $m_2 = -1/2)$ of the same operators on the spin-$1/2$ representation space. 
Therefore, it follows from the Clebsch-Gordan decomposition that
\begin{equation}
\begin{split}
      \ket{D_n^i}\otimes (\gamma \ket{0} + \delta \ket{1}) &= \ket{j_1, m_1}\otimes (\gamma \ket{1/2, 1/2} + \delta \ket{1/2, -1/2}) \\
      &= \gamma \sqrt{\frac{1}{2}\left(1 + \frac{(n+1)/2 - i}{(n+1)/2}\right)}\ket{j_1 + 1/2, m_1 + 1/2}\\ & - \gamma \sqrt{\frac{1}{2}\left(1 - \frac{(n+1)/2 - i}{(n+1)/2}\right)}\ket{j_1 - 1/2, m_1 + 1/2}\\
      &+ \delta \sqrt{\frac{1}{2}\left(1 - \frac{(n-1)/2 - i}{(n+1)/2}\right)}\ket{j_1 + 1/2, m_1 - 1/2}\\ & +\delta \sqrt{\frac{1}{2}\left(1 + \frac{(n-1)/2 - i}{(n+1)/2}\right)}\ket{j_1 - 1/2, m_1 - 1/2}
\end{split}
\end{equation}
After adding a qubit and applying a gate $U(\gamma, \delta^*)$, the resulting intermediate state is thus a superposition of states with total angular momentum $j_1 - 1/2$ or $j_1 + 1/2$. It follows again from  standard angular momentum theory that the states $\ket{j_1 + 1/2, m_1 \pm 1/2}$ associated to a total angular momentum of $j_1 + 1/2$ can be identified as Dicke states  defined on $n+1$ qubits that are therefore invariant under any permutation of all these qubits,
\begin{equation}
    \ket{j_1 + 1/2, m_1 + 1/2} = \ket{D_{n+1}^{i}}, \quad \ket{j_1 + 1/2, m_1 - 1/2} = \ket{D_{n+1}^{i+1}}.
\end{equation}
%By applying Schur-Weyl duality, it follows that the representation of the symmetric group, arising from the action of qubit permutations on the subspace of states with total angular momentum $j_1 - 1/2$, is both irreducible and isomorphic to the standard representation of the symmetric group. In particular, there is no state in this subspace that is invariant under any permutation of the qubits. Therefore, the task of realizing a measurement to check wheter we are in a symmetric state or not amounts to to a measurement of the total angular momentum. The protocol is a success if we measure $j_1 + 1/2$ and fails otherwise. 
The subsequent step in Protocol 2 involves measuring the total angular momentum to determine whether the observed state has angular momentum $j_1 + 1/2$ or $j_1 - 1/2$. For certain physical implementations of qubits, direct measurements of the total angular momentum may be accessible, making this step straightforward. In implementations where such measurements are unavailable, and only qubit measurements in the 
$\{\ket{0}, \ket{1}\}$ basis are possible, it is still feasible to differentiate between the states associated with total angular momentum $j_1 + 1/2$ and $j_1 - 1/2$ by using a circuit that combines SWAP gates with the quantum phase estimation algorithm (QPE). Some details on the implementation of this approach are presented in Appendix \ref{App:B}. Either way, if the measurement confirms a maximal angular momentum $j_1 + 1/2$, the protocol results in the following transformation of Dicke states
\begin{equation}
\begin{split}
      \ket{D_n^i}\otimes (\gamma \ket{0} + \delta \ket{1}) &\longrightarrow \gamma \sqrt{\frac{n+1 -i}{n+1}}\ket{D_{n+1}^i}+ \delta \sqrt{\frac{i+1}{n+1}}\ket{D_{n+1}^{i+1}}.
\end{split}
\end{equation}
In particular, a success of the protocol corresponds to the following transformation of the inital $n$ qubit state $\ket{\psi}$ into the state $\ket{\psi'}$:
\begin{equation}
    \ket{\psi} \longrightarrow  \ket{\psi'} = \kappa \sum_{i =0}^{n+1} \psi'_{n+1,i}  \ket{D_{n+1}^i}
\end{equation}
with $\kappa$ a normalization constant and with the coefficients $\psi'_{n+1,i}$  given as follows in terms of the coefficients $\psi_{n,i}$ defining the initial symmetric state $\ket{\psi}$
\begin{equation}
     \psi_{n+1,i} =  \gamma \sqrt{{n+1 -i}}\psi_{n,i}+ \delta \sqrt{{i}}\psi_{n,i-1}.
\end{equation}
The transformation induced by a successful application of Protocol 2 is thus given, up to a normalization constant $\kappa$, by the linear operator $P_2(\gamma, \delta): \mathcal{D} \to \mathcal{D}$, whose action on a given Dicke state is
\begin{equation}
   P_2(\gamma, \delta)  \ket{D_{n}^i} = \gamma \sqrt{{n+1 -i}}\ket{D_{n+1}^i}+ \delta \sqrt{{i+1}}\ket{D_{n+1}^{i+1}}.
\end{equation}

\subsection{Combination of the Protocols 1 and 2}

In the previous subsection, we have introduced operators $P_1(\alpha, \beta)$ and $P_2(\gamma,\delta)$ which implement the transformations induced by Protocols 1 and 2 when successful. If both protocols are successfully applied in succession on a symmetric state $\ket{\psi}$ defined on $n$ qubits, it is straightforward to see that their composition preserves the number $n$ of qubits and results (up to a normalization constant) in the following transformation:
\begin{equation}\label{p1p2psi}
    P_1(\alpha, \beta) P_2(\gamma,\delta)  \ket{\psi} = \sum_{i = 0}^n \hat{\psi}_{n,i}\ket{D_n^i},
\end{equation}
with $\hat{\psi}_{n,i}$ expressed in terms of the defining coefficient $\psi_{n,i}$ of $\ket{\psi}$ as,
\begin{equation}\label{eq:actp1p2}
    \hat{\psi}_{n,i}   =  \gamma  \beta \sqrt{(n -i)(i+1)}\psi_{n,i+1} + \left( \alpha \gamma (n-i) + \delta \beta i\right)\psi_{n,i} + \alpha \delta \sqrt{i(n-i+1)}\psi_{n,i-1}.
\end{equation}
This corresponds to the case where the Protocol 2 is applied first. Inverting the order of operations, one similarly finds that the transformation induced by first applying Protocol 1 successfully, followed by a successful application of Protocol 2, corresponds (up to a normalization constant) to:
\begin{equation}
 P_2(\gamma,\delta)   P_1(\alpha, \beta)   \ket{\psi} = \sum_{i = 0}^n  \tilde{\psi}_{n,i}\ket{D_n^i},
\end{equation}
with $\tilde{\psi}_{n,i}$ now given by
\begin{equation}
    \tilde{\psi}_{n,i}   =  \gamma  \beta \sqrt{(n -i)(i+1)}\psi_{n,i+1} + \left( \alpha \gamma (n-i+1) + \delta \beta( i + 1)\right)\psi_{n,i} + \alpha \delta \sqrt{i(n-i+1)}\psi_{n,i-1}.
\end{equation}
Next, we apply these results, along with the explicit actions of $P_1(\alpha,\beta)$ and $P_2(\gamma,\delta)$ to obtain an algebraic description of the set of transformations induced by Protocols $1$ and $2$.

\section{Algebra of protocol-induced transformations}\label{sec:5}

In this section, we demonstrate that the operators $P_1(\alpha,\beta)$ and $P_2(\gamma,\delta)$ provide a representation of the direct sum of two copies of the Weyl algebra that form the dynamical algebra associated to Protocols $1$ and $2$.

\subsection{Dynamical algebra}

Denote by $a_1$, $a_1^\dagger$,  $a_2$ and $a_2^\dagger$ the operators on $\mathcal{D}$ that are defined as follows by their actions on Dicke states: 
\begin{equation}
    a_1 \ket{D_n^i} = \sqrt{n-i} \ket{D_{n-1}^i}, \quad  a_1^\dagger \ket{D_n^i} = \sqrt{n+1-i} \ket{D_{n+1}^i},
\end{equation}
\begin{equation}
    a_2 \ket{D_n^i} = \sqrt{i} \ket{D_{n-1}^{i-1}}, \quad  a_2^\dagger  \ket{D_n^i} = \sqrt{i+1} \ket{D_{n+1}^{i+1}}.
\end{equation}
It is straightforward to verify that each pair $(a_i,a_i^\dagger) \; i=1,2$, satisfies the defining commutation relation of the Weyl algebra $W(1)$ and commutes with the operators from the other pair, i.e.
\begin{equation}\label{eq:2weyl}
    [a_i, a_j^\dagger] =  \delta_{ij}, \quad i,j \in \{1,2\}.
\end{equation}
 The connection to the algebra generated by the operators $P_1(\alpha, \beta)$ and $P_2(\gamma, \delta)$ is established by observing that they can be expressed as linear combinations of the generators $a_i$ and $a_i^\dagger$:
\begin{equation}
    P_1(\alpha,\beta) = \alpha a_1 + \beta a_2, \quad  P_2(\gamma,\delta) = \gamma a_1^\dagger + \delta a_2^\dagger.
\end{equation}
This casts the Weyl algebra $W(2)=W(1)\oplus W(1) $ as the dynamical algebra of Dicke state transformations associated to the Protocols 1 and 2.
At this stage, two key observations are worth making. First, the operators $P_1(\alpha,\beta)$ and $P_2(\gamma,\delta)$ generate the entire algebra spanned by the operators $a_i$ and $a_i^\dagger$, as evidenced by the following identifications:
\begin{equation}
     a_1 =P_1(1,0), \quad a_1^\dagger  = P_2(1,0), \quad  a_2=P_1(0,1), \quad a_2^\dagger  = P_2(0,1).
\end{equation}
Next, the operator $ \mathcal{N}$ which acts diagonally on the Dicke states and measures the number of qubits, i.e. $\mathcal{N} \ket{D_n^i} =  n \ket{D_n^i}$,
can also be embedded in (the envelopping algebra of) W(2) as
\begin{equation}\label{eq:2dQAO}
    \mathcal{N} = a_1^\dagger a_1 + a_2^\dagger a_2.
\end{equation}
Using the relation \eqref{eq:2weyl}, it is straightforward to compute the commutation relations between the operators $P_1(\alpha,\beta)$, $P_2(\gamma,\delta)$ and $\mathcal{N}$ to find:
\begin{equation} [P_1(\alpha,\beta),P_1(\alpha',\beta')] = [P_2(\gamma, \delta),P_2(\gamma',\delta')] = 0,
\end{equation}
\begin{equation}\label{eq:reldyn}
[\mathcal{N},P_2(\gamma,\delta) ] = P_2(\gamma,\delta), \quad [\mathcal{N},P_1(\alpha,\beta) ] = -P_1(\alpha,\beta), \quad [P_2(\gamma,\delta),P_1(\alpha,\beta)] = \alpha\gamma + \delta \beta.
\end{equation}
In particular, we observe that for a fixed choice of single-qubit gates $U(\alpha, \beta)$ and $U(\gamma, \delta^*)$, the algebra generated by the transformations $P_1(\alpha, \beta)$, $P_2(\gamma, \delta)$, and the operator $\mathcal{N}$ is isomorphic to a single copy of the Weyl algebra $W(1)$, provided that
\begin{equation}
     \bra{0} U(\alpha,\beta)U(\gamma,\delta^*) \ket{0} = \alpha\gamma + \delta \beta \neq 0 .
\end{equation}

\subsection{Symmetry algebra}

The operator $\mathcal{N}$ is degenerate: the $n+1$ Dicke vectors $\ket{D_n^i}$ with $i = 0, 1, \dots, n$ span the eigenspace associated with its eigenvalue $n$. We now wish to characterize the algebra of operators, based on Protocols 1 and 2, which commute with $\mathcal{N}$ and provide maps between symmetric states defined on $n$ qubits. It is obvious that $\mathcal{N}$ commutes with $P_2(\gamma,\delta)P_1(\alpha,\beta)$ and $P_1(\alpha,\beta) P_2(\gamma,\delta)$,
\begin{equation}
    [\mathcal{N}, P_1(\alpha,\beta)P_2(\gamma,\delta)] =  [\mathcal{N}, P_2(\gamma,\delta)P_1(\alpha,\beta)] = 0.
\end{equation}
Given relations \eqref{eq:reldyn}, it is immediate to check that these two operators i.e. $P_1(\alpha,\beta) P_2(\gamma,\delta)$ and $P_2(\gamma,\delta)P_1(\alpha,\beta)$, only differ by a constant so that we can restrict our attention to $P_1(\alpha,\beta)P_2(\gamma,\delta)$. We shall thus consider the algebra generated by the operators $P_1(\alpha,\beta)P_2(\gamma,\delta)$ for different choices of the parameters $\alpha,\beta$ and $\gamma,\delta$ (determining the unitary one-qubit gates in the protocols). 

We have already indicated in Section \ref{sec:3} that the Dicke states span a $(n+1)$-dimensional irreducible submodule of the $n$-fold tensor product of the spin-$\frac{1}{2}$ representation of $\mathfrak{su}(2)$. We observed that the adjacency matrix $A$ of the hypercube and its dual $A^*$ are respectively given by the matrices $J^x$ and $J^z$ representing (see \eqref{eq:defJ}) the abstract generators $j^x$ and $j^z$ of $\mathfrak{su}(2)$ in $(\frac{1}{2})^{\otimes n}$. Precisely, $A=2J^x$ and $A^*=2J^z$. The restrictions of $J^x$ and $J^z$ on the span of the Dicke states $|D_n^{i}\rangle \; i=0,\dots,n$, is hence provided already by the formulas \eqref{AonD} and \eqref{A*onD} giving the actions of $A$ and $A^*$ on those states.

%Equation \eqref{eq:2dQAO} demonstrates that $\mathcal{N}$ shares the same embedding into two copies of the Weyl algebra as the Hamiltonian of the two-dimensional isotropic quantum harmonic oscillator. Consequently, the symmetry algebra arising from the combination of Protocols $1$ and $2$ is expected to be somehow related to $\mathfrak{su}(2)$, which is the symmetry algebra of this simple quantum mechanical system. 

%Indeed, recall that the $\mathfrak{su}(2)$ representation resulting from the composition of $n$ spin-$1/2$ representations is expressed as
%\begin{equation}\label{eq:defj}
  %  j^\ell = \sum_{i=1}^n \underbrace{I\otimes...\otimes I}_{i-1\text{ times}}\otimes\ \frac{\sigma_\ell}{2}\otimes \underbrace{I\otimes...\otimes I}_{N-i\text{ times}}, \quad \ell = x,y,z.
%\end{equation}
%\textcolor{blue}{(Recall that the adjacency matrix $A$ of the hypercube and its dual $A^*$ had been identified in \eqref{Adj} and \eqref{Adjd} as $A=2j^x$ and $A^*=2j^z$ in this representation.)}
%\textcolor{blue}{As explained in Section 3}This representation is reducible and possesses a unique irreducible $n+1$-dimensional submodule, corresponding to the space $\mathcal{D}_n$, which is spanned by the Dicke states defined on $n$ qubits. The action of the generators $j^\ell$ on these Dicke states can be determined using a combinatorial argument and is expressed as follows:
%\begin{equation}
  %  j^x \ket{D_n^i}  = \frac{\sqrt{i(n -i+1)}}{2} \ket{D_n^{i-1}}  + \frac{\sqrt{(i+1)(n-i)}}{2}\ket{D_n^{i+1}}
%\end{equation}

It is convenient to supplement these with
\begin{equation} \label{Jy}
    J^y \ket{D_n^i}  = i[J^x,J^z] \ket{D_n^i}= -i\frac{\sqrt{i(n -i+1)}}{2} \ket{D_n^{i-1}}  + i \frac{\sqrt{(i+1)(n-i)}}{2}\ket{D_n^{i+1}}
\end{equation}
to equip the reader with the explicit formulas giving $J^{a}|D_n^{i}\rangle$ for all $a=x,y,z$. From equations \eqref{AonD}, \eqref{A*onD} and \eqref{Jy}, we can infer from \eqref{eq:actp1p2} the following identification of the operators $P_1(\alpha, \beta)P_2(\gamma, \delta)$ in terms of the operators $J^{a}, \; a=x,y,z, \;\text{and} \; \mathcal{N}$ restricted to the space $\mathcal{D}_n$, thereby establishing a connection between $\mathfrak{su}(2)$ and the symmetry algebra of $\mathcal{N}$:
\begin{equation}\label{eq:idf1}
P_1(\alpha,\beta)P_2(\gamma,\delta)\big|_{\mathcal{D}_n}  = \left(v_x {J}^x + v_y {J}^y + v_z {J}^z + v_0 \mathcal{N}\right)\big|_{\mathcal{D}_n},
\end{equation}
with
\begin{equation}
    \gamma \beta = \frac{v_x - i v_y}{2} , \quad \alpha \delta = \frac{v_x + i v_y}{2} ,\quad   -\delta \beta + \alpha \gamma = v_z, \quad \alpha \gamma = \frac{1}{2} v_z + v_0,
\end{equation}
or equivalently
\begin{equation}
    v_x = \alpha \delta + \gamma \beta, \quad v_y = i (\gamma \beta - \alpha \delta), \quad v_z =  -\delta \beta + \alpha \gamma, \quad v_0 = \frac{1}{2} (\alpha \gamma  +  \delta \beta) .
\end{equation}
Since the coefficients $v_x$, $v_y$, $v_z$, and $v_0$ are generally complex, it follows that the algebra generated by the composition of Protocols $1$ and $2$, which constitutes the symmetry algebra of the operator measuring the number of qubits $\mathcal{N}$, is in fact the complexification of $\mathfrak{su}(2)$ (with $\mathcal{N}$ related to the Casimir element $\boldsymbol{j}^2 \rightarrow \frac{\mathcal{N}}{2}\left(\frac{\mathcal{N}+1}{2}\right)$).

\subsection{Diagonalization of $P_1(\alpha,\beta)P_2(\gamma,\delta)$ and Krawtchouk polynomials}

For a suitable choice of $\alpha$, $\beta$, $\gamma$, and $\delta$ such that $P_1(\alpha,\beta)P_2(\gamma,\delta)$ is non-degenerate and diagonalizable, the common eigenbases of $\mathcal{N}$ and $P_1(\alpha,\beta)P_2(\gamma,\delta)$ form a basis for the space $\mathcal{D}$ of symmetric states. This basis is particularly noteworthy, as it consists of fixed points of transformations induced by the composition of Protocols $1$ and $2$. In this subsection, we therefore focus on the diagonalization of the operator $P_1(\alpha,\beta)P_2(\gamma,\delta)$ on each eigenspace $\mathcal{D}_n$ of $\mathcal{N}$. Let $\phi$ and $\theta$ be complex parameters defined by
\begin{equation}
    \tan\theta = \frac{v_y}{v_x} = \frac{i(\gamma \beta -\alpha\delta)}{(\alpha \delta + \beta\gamma)}, \quad \tan \phi = - \frac{\sqrt{v_x^2 + v_y^2}}{v_z} = -\frac{2\sqrt{\alpha \beta \gamma \delta}}{\alpha \gamma - \delta \beta}, \quad \cos \phi = \frac{\alpha \gamma - \beta \delta}{\alpha \gamma + \delta \beta}.
\end{equation}
%\begin{equation}
%   \frac{\alpha \delta}{\beta \gamma} = e^{2 i \theta}, \quad \tan\phi =\frac{2 \alpha \delta e^{-i\theta}}{\alpha \gamma - \delta \beta}.
%\end{equation}
They allow the introduction of the $(n+1) \times (n+1)$ matrix $B$ defined as follows in terms of the matrices $J^a|_{\mathcal{D}_n}$, hereafter denoted simply by ${J}^{a}$ (with the restriction to ${\mathcal{D}_n}$ understood in this subsection) to lighten the notation,
\begin{equation}
    B = e^{i\theta J^z} e^{i\phi J^y}.
\end{equation}
Using the Baker–Campbell–Hausdorff formula, the identification \eqref{eq:idf1} and the commutation relations of the generators of $\mathfrak{su}(2)$, we find that the change of basis associated to $B$ diagonalizes the operator $P_1(\alpha,\beta)P_2(\gamma,\delta) $, i.e.
\begin{equation}
    B^{-1} P_1(\alpha,\beta)P_2(\gamma,\delta) B = {(\alpha \gamma + \delta \beta)}\left(J^z  + \frac{\mathcal{N}}{2}\right),
\end{equation}
where we used $P_1(\alpha,\beta)P_2(\gamma,\delta) = P_1(\alpha,\beta)P_2(\gamma,\delta)|_{\mathcal{D}_n}$ and $\mathcal{N}=\mathcal{N}|_{\mathcal{D}_n}= n$ to keep the notation simple. We thus find that the vectors $B \ket{D_{n}^i}$, with $ n = 0,1,2, \dots$ and $i = 0,1, \dots, n$, provide a basis of $\mathcal{D}$ consisting of fixed points of the composition of Protocols $1$ and $2$. In other words,
\begin{equation}
    \mathcal{D} = \text{span}\{B\ket{D_{n}^i}\, | \, (i,n) \in \mathbb{N}^2, i\leq n\},
\end{equation}
with
\begin{equation}
    P_1(\alpha,\beta)P_2(\gamma,\delta) B\ket{D_{n}^i} = \lambda_i B\ket{D_{n}^i}, \quad   \lambda_i = \left(\alpha \gamma +  \beta \delta\right)\left(n - i\right).
\end{equation}
The decomposition of the fixed points $B\ket{D_{n}^j}$ in the original basis of Dicke states $\ket{D_{n}^i}$ can be computed straightforwardly by leveraging a three-term recurrence relation in a way quite similar to how the diagonalization of the adjacency matrix of the hypercube was carried out in Section \ref{sec:3}. This relation is obtained by evaluating the coefficient $\bra{D_{n}^i}P_1(\alpha,\beta)P_2(\gamma,\delta) B\ket{D_{n}^j}$ using both the right and left actions to find: 
\begin{equation}\label{eq:3tr}
    \lambda_j \psi_{n,i}   =  \gamma  \beta \sqrt{(n -i)(i+1)}\psi_{n,i+1} + \left( \alpha \gamma (n-i) + \delta \beta i\right)\psi_{n,i} + \alpha \delta \sqrt{i(n-i+1)}\psi_{n,i-1},
\end{equation}
where $\psi_{n,i} := \bra{D_{n}^j} B\ket{D_{n}^i}$. This relation is easily seen to be transformable into the three term recurrence relation of Krawtchouk polynomials and leads to 
\begin{equation}
    B\ket{D_{n}^j} =  \sum_{i = 0}^{n} \psi_{n,i} \ket{D_{n}^i}, \quad \psi_{n,i} = \psi_{n,0}\left(\frac{\delta}{\gamma}\right)^i \sqrt{\binom{n}{i} }K_i\left(j; \frac{\beta\delta}{\alpha \gamma + \beta\delta} , n\right).
\end{equation}

For some choice of the parameters $\alpha$, $\beta$, $\gamma$ and $\delta$, the operator $P_1(\alpha,\beta)P_2(\gamma,\delta) $ is proportional to a hermitian operator, with $\theta$ and $\phi$ both real and $B$ is furthermore a unitary transformation. In such instances, the fixed point states $B \ket{D_n^{i}}$ can  be constructed from the Dicke states $\ket{D_n^{i}}$ by the application of multiple copies of a single qubit gate,
\begin{equation}\label{eq:Bunit}
    B = ( - i\,  U(\mu, \nu))^{\otimes n}, \quad \mu = i  e^{ i \frac{\theta}{2}} \cos \frac{\phi}{2}, \quad \nu=  i e^{ i \frac{\theta}{2}} \sin \frac{\phi}{2}.
\end{equation}
As an example, consider the case where $\alpha = \beta = \gamma = \delta = 1/\sqrt{2}$. This specific scenario arises when both Protocols $1$ and $2$ use the Hadamard gate as the single-qubit operation. In this case, the transformation induced by the composition of the two protocols is equivalent to the action of $J^x + \mathcal{N}/2$ on Dicke states. Although this action is not diagonal, it was shown in Section \ref{sec:3} that the corresponding matrix can be diagonalized using the Hadamard transform. This result is consistent with \eqref{eq:Bunit}, which gives $B = H^{\otimes n}$. Consequently, a state invariant under the composition of the two protocols, with $\alpha = \beta = \gamma = \delta = 1/\sqrt{2}$, can be prepared by applying the Hadamard transform to a Dicke state.
%with $\psi_{n,0}$ a constant term.
%\begin{equation}
%\psi_{n,0}=\left(\frac{\alpha \gamma }{\alpha \gamma  + \beta \delta}\right)^{n/2}\left(\frac{ \beta \gamma}{\alpha \delta }\right)^{n/4} \left(\frac{\delta \beta}{\alpha\gamma}\right)^{j/2} \sqrt{\binom{n}{j}}
%\end{equation}
\section{Some applications}\label{sec:6}

We introduced two protocols and provided an algebraic description for the set of transformations they induce. In this section, we explore potential applications of this framework, focusing on the preparation of Dicke states and the characterization of the states obtained after multiple iterations of the two protocols.

\subsection{Preparation of symmetric states}

Let $\ket{\psi}$ corresponds to an arbitrary symmetric state on $n$ qubits, with overlaps $\bra{D_n^i}\ket{\psi} = \psi_{n,i}$. We are interested in the preparation of this state using a device allowing the realization of Protocol $2$. In particular, we are wondering how the state $\ket{\psi}$ can be constructed starting from the vacuum state $\ket{D_0^0}$ that has no qubits. This state has the property of being anihilated by the transformations $a_1$ and $a_2$ that can be induced by the Protocol $1$:

%we are interested in its preparation from the state associated to the absence of any qubits is symmetric and denoted $\ket{D_0^0}$. This state also corresponds to the unique vacuum state associated to the representation of two copies of the Weyl group given by the transformation induced by Protocols $1$ and $2$,
\begin{equation}
    a_1 \ket{D_0^0} = 0, \quad a_2 \ket{D_0^0} =0.
\end{equation}
All Dicke states can be obtained from repeated applications of the creation operators $a_1^\dagger$ and $a_2^\dagger$, and are the analogs of Fock space basis vectors,
\begin{equation}\label{eq:fock}
    \ket{D_n^i} = \frac{1}{\sqrt{i! (n-i)! }}\left(a_2^\dagger\right)^{i} \left(a_1^\dagger\right)^{n-i} \ket{D_0^0}.
\end{equation}
Now let us define $x_i$, $i = 1,2, \dots n$ as the zeros of the polynomial $Q(x)$ which is a generating function for the overlaps $\psi_{n,i}$,
\begin{equation}
    Q(x) = \sum_{i=0}^n \frac{\psi_{n,i} }{\sqrt{i! (n-i)!}} x^i,
\end{equation}
 and introduce the parameters $\gamma_i$ and $\delta_i$ related as follows to the roots $x_i$ of $Q(x)$,
\begin{equation}
    x_i = -\frac{\gamma_i}{\delta_i}, \quad |\gamma_i|^2 + |\delta_i|^2 = 1, \quad i = 1,2,\dots n.
\end{equation}
Using equation \eqref{eq:fock}, it is a straightforward to show that the state $\ket{\psi}$ can be constructed from the vacuum through multiple successful applications of Protocol $2$,
\begin{equation}
    \ket{\psi} \propto \prod_{i = 1}^n P_2(\gamma_i, \delta_i) \ket{D_0^0},
\end{equation}
which is the answer to the question we asked at the beginning of this subsection.

\subsection{Iterated application of the two protocols}

Consider now the scenario where both protocols are alternated repeatedly with fixed parameters $\alpha$, $\beta$, $\gamma$ and $\delta$. Given that they are successful, one might be interested in determining the asymptotic transformation of an arbitary initial symmetric state $\ket{\psi}$. In the case where $P_1(\alpha,\beta) P_2(\gamma, \delta)$ admits an eigenbasis, one finds that any symmetric initial state $\ket{\psi}$ on $n$ qubits can be decomposed as
\begin{equation}
   \ket{\psi} = \sum_{i =0}^n \chi_{i} B \ket{D_n^i}, \quad \chi_i \in \mathbb{C}.
\end{equation}
Under $N$ application of both protocols, one thus gets
\begin{equation}
   (P_1(\alpha,\beta) P_2(\gamma, \delta))^N\ket{\psi} = \sum_{i =0}^n (\alpha \gamma + \beta \delta)^N(n-i)^N\chi_{i} B \ket{D_n^i}.
\end{equation}
For $N$ large and assuming that $k$ is the smallest integer such that the expansion parameter $\chi_{k} \neq 0$, we find that
\begin{equation}
     (P_1(\alpha,\beta) P_2(\gamma, \delta))^N\ket{\psi} = (\alpha \gamma + \beta \delta)^N(n-k)^N \left(\chi_{k} B \ket{D_n^k} + \left(\frac{n-k-1}{n-k}\right)^N\chi_{k+1} B \ket{D_n^{k+1}} + ... \right)
\end{equation}
Thus, up to exponentially smaller subleading terms, we find that
\begin{equation}
    (P_1(\alpha,\beta) P_2(\gamma, \delta))^N\ket{\psi} \approx (\alpha \gamma + \beta \delta)^N(n-k)^N \chi_{k} B \ket{D_n^k}, \quad N \rightarrow \infty.
\end{equation}

It is understood that the probability to get this result becomes asymptotically small; the point here is to indicate to what the state tends as the combination of the two protocols is repeated.

\section{Conclusion}\label{sec:7}

We have considered the set of states of $n$ qubits that are totally symmetric under permutations, along with their transformations as induced by two simple protocols. The first protocol involves the measurement and removal of a qubit, while the second adds a qubit in combination with a measurement that symmetrizes the total $(n+1)$-qubit state. Using these protocols as building blocks, we found that the transformations they induce on the space of totally symmetric states realize a representation of the Weyl algebra $W(2)$. The upshot is that this dynamical algebra of transformations on the span of Dicke states can be engendered by the protocols.

We demonstrated that the composition of the two protocols induces transformations that preserve the number of qubits and provides a representation of the generators of (a complexification of) $\mathfrak{su}(2)$. The fixed point vector states associated with the combination of the protocols were identified and shown to have Krawtchouk polynomials as expansion coefficients over Dicke states. Finally, we discussed potential applications of the framework. By adopting a formalism akin to the second quantization approach, we showed that Protocol~2 allows the construction on the vacuum state of any totally symmetric state on $n$ qubits with $n$ successful applications of Protocol~2. Additionally, we observed that the fixed points of the transformations induced by the combination of Protocols~1 and~2 can be interpreted as the asymptotic states obtained after a large number of repetitions of both protocols on an arbitrary state.

This algebraic description informs the possible manipulations of Dicke states that can be achieved from the use of the two simple protocols that have been considered. More elaborate protocols should be envisaged and could lead to more refined and elaborate dynamical algebras.

Some $q$-analogs of the Dicke states have been introduced \cite{li2015entanglement}. The $q$-Dicke states  are rooted in the representation theory of $U_q(\mathfrak{sl}_2)$. Their combinatorial interpretation has been studied recently \cite{bernard2024q} and has shown that these states enjoy a relation with a weighted hypercube much similar to the one described here between Dicke states and the regular hypercube. Of interest also are the qudit Dicke states as well as their $q$ deformations which relate to higher rank (quantum) algebras \cite{raveh2024q}. In future work, it would be interesting to investigate whether the protocol-based framework developed in this paper could be extended to situations that feature these generalized Dicke states.

\appendix

\section{ Krawtchouk polynomials}\label{app:A}

We reproduce for easy reference the definition and three term recurrence relation of the Krawtchouk polynomials $K_i(x,p,n)$ \cite{koekoek2010hypergeometric}.
These polynomials are defined as a terminating hypergeometric series:

\begin{equation}
K_i(x;p,n) = \hg{2}{1}\argu{-i, -x}{-n}{\frac{1}{p}}, i=0,1,\dots,n.
\end{equation}
It is manifest from their definition that the Krawtchouk polynomials are self-dual, that is:
\begin{equation}
 K_i(x;p,n)=K_x(i;p,n), \quad x,i\in {0,1, \dots,n} .  
\end{equation}
They obey the three term recurrence relation:
\begin{equation}
    -xK_i(x;p,n) =p(n-i)K_{i+1}(x;p,n) -[p(n-i)+i(1-p)]K_i(x;p,n) + i(1-p)K_{i-1}(x;p,n).
\end{equation}

\section{Circuit for total angular momentum measurement}\label{App:B}

Protocol 2 requires the measurement of the total angular momentum of $n+1$ qubits, assuming they are in a superposition of states with maximal total angular momentum $j_1 + \frac{1}{2} = \frac{n+1}{2}$ and of states with total angular momentum $j_1 - \frac{1}{2}$.

From Schur-Weyl duality, it follows that for a given eigenvalue equal to $m_1+\frac{1}{2}$ or $m_1-\frac{1}{2}$ of the $z$-component of the total angular momentum operator, there exists a unique vector $\ket{j_1 + \frac{1}{2}, m_1 \pm \frac{1}{2}}$ that is invariant under any permutation of qubits. Similarly, the same duality implies the existence of an $n$-dimensional subspace associated with $j_1 - \frac{1}{2}$ and $m_1 \pm \frac{1}{2}$, where the operators representing qubit permutations act according to the standard irreducible representation of the symmetric group $S_{n+1}$.

Let $\sigma$ denote the matrix corresponding to the cyclic permutation of the $n+1$ qubits. From the previous remarks and the eigenvalues of $\sigma$ in the trivial and standard representations of $S_{n+1}$, we have that:
\begin{equation}
    \sigma \ket{j_1 + 1/2, m_1 \pm  1/2} = \ket{j_1 +  1/2, m_1 \pm  1/2},
\end{equation}
and we know that there exists a basis $\{\ket{j_1 - \frac{1}{2}, m_1 \pm \frac{1}{2}, \ell} \mid \ell = 1, 2, \dots, n\}$ for the subspace associated with $j_1 - \frac{1}{2}$ and $m_1 \pm \frac{1}{2}$, such that:
\begin{equation}
    \sigma \ket{j_1 -  1/2, m_1 \pm  1/2, \ell} = e^{\frac{2i\pi \ell}{n+1}} \ket{j_1 - 1/2, m_1 \pm  1/2, \ell}.
\end{equation}

To distinguish between the states with $j_1 + \frac{1}{2}$ and $j_1 - \frac{1}{2}$, one can employ the quantum phase estimation (QPE) algorithm, with $\sigma$ serving as the unitary operator. Indeed, the outcome of the algorithm will be a measurement of the phase induced by $\sigma$ on our state. Therefore, if the phase is found to be zero, the output state is in the eigenspace of $\sigma$ with eigenvalue $1$, and thus has total angular momentum $j_1 + 1/2$. Otherwise, we know that we are in an eigenspace of $\sigma$ associated to the standard representation and a total angular momentum of $j_1 - 1/2$. In that case, Protocols $2$ fails and the state is discarded. 

Note that, since $\sigma$ corresponds to the cyclic permutation of qubits, it can be implemented using SWAP gates. The controlled version of $\sigma$ (required for QPE) can then be realized by replacing the SWAP gates with Fredkin gates.

\section*{Acknowledgments}
The authors gratefully acknowledge stimulating discussions with Eddie Farhi.
PAB holds an Alexander-Graham-Bell scholarship from the Natural Sciences and Engineering Research Council (NSERC) of Canada. LV is funded in part through a discovery grant from NSERC.

\providecommand{\href}[2]{#2}\begingroup\raggedright \endgroup


\begin{thebibliography}{10}

\bibitem{dicke1954coherence}
R.~H. Dicke, ``Coherence in spontaneous radiation processes,'' \href{http://dx.doi.org/https://doi.org/10.1103/PhysRev.93.99}{{\em Physical review} {\bfseries 93}, 99 (1954)}.

\bibitem{farhi2014quantum}
E.~Farhi, J.~Goldstone, and S.~Gutmann, ``A quantum approximate optimization algorithm,'' \href{http://dx.doi.org/https://doi.org/10.48550/arXiv.1411.4028}{{\em arXiv preprint arXiv:1411.4028}  (2014)}.

\bibitem{childs2000finding}
A.~M. Childs, E.~Farhi, J.~Goldstone, and S.~Gutmann, ``{Finding cliques by quantum adiabatic evolution},'' \href{http://dx.doi.org/https://doi.org/10.48550/arXiv.quant-ph/0012104}{{\em arXiv preprint quant-ph/0012104}  (2000)}.

\bibitem{bartschi2019deterministic}
A.~B{\"a}rtschi and S.~Eidenbenz, \href{http://dx.doi.org/https://doi.org/10.1007/978-3-030-25027-0_9}{``{Deterministic preparation of Dicke states},''} in {\em International Symposium on Fundamentals of Computation Theory}, pp.~126--139, Springer.
\newblock 2019.

\bibitem{buhrman2024state}
H.~Buhrman, M.~Folkertsma, B.~Loff, and N.~M. Neumann, ``State preparation by shallow circuits using feed forward,'' \href{http://dx.doi.org/https://doi.org/10.22331/q-2024-12-09-1552}{{\em Quantum} {\bfseries 8}, 1552 (2024)}.

\bibitem{piroli2024approximating}
L.~Piroli, G.~Styliaris, and J.~I. Cirac, ``Approximating many-body quantum states with quantum circuits and measurements,'' \href{http://dx.doi.org/https://doi.org/10.1103/PhysRevLett.133.230401}{{\em Physical Review Letters} {\bfseries 133}, 230401 (2024)}.

\bibitem{nepomechie2023qudit}
R.~I. Nepomechie and D.~Raveh, ``{Qudit Dicke state preparation},'' \href{http://dx.doi.org/https://doi.org/10.48550/arXiv.2301.04989}{{\em arXiv preprint arXiv:2301.04989}  (2023)}.

\bibitem{yu2024efficient}
J.~Yu, S.~R. Muleady, Y.-X. Wang, N.~Schine, A.~V. Gorshkov, and A.~M. Childs, ``{Efficient preparation of Dicke states},'' \href{http://dx.doi.org/https://doi.org/10.48550/arXiv.2411.03428}{{\em arXiv preprint arXiv:2411.03428}  (2024)}.

\bibitem{liu2024low}
Z.~Liu, A.~M. Childs, and D.~Gottesman, ``{Low-depth quantum symmetrization},'' \href{http://dx.doi.org/https://doi.org/10.48550/arXiv.2411.04019}{{\em arXiv preprint arXiv:2411.04019}  (2024)}.

\bibitem{kobayashi2014universal}
T.~Kobayashi, R.~Ikuta, {\c{S}}.~K. {\"O}zdemir, M.~Tame, T.~Yamamoto, M.~Koashi, and N.~Imoto, ``{Universal gates for transforming multipartite entangled Dicke states},'' \href{http://dx.doi.org/10.1088/1367-2630/16/2/023005}{{\em New Journal of Physics} {\bfseries 16}, 023005 (2014)}.

\bibitem{bernard2024q}
P.-A. Bernard, {\'E}.~Poliquin, and L.~Vinet, ``{A $q$-version of the relation between the hypercube, the Krawtchouk chain and Dicke states},'' \href{http://dx.doi.org/10.1088/1751-8121/ad94b9}{{\em Journal of Physics A: Mathematical and Theoretical}  (2024)}.

\bibitem{bernard2018graph}
P.-A. Bernard, A.~Chan, {\'E}.~Loranger, C.~Tamon, and L.~Vinet, ``A graph with fractional revival,'' \href{http://dx.doi.org/https://doi.org/10.1016/j.physleta.2017.12.001}{{\em Physics Letters A} {\bfseries 382}, 259--264 (2018)}.

\bibitem{go2002terwilliger}
J.~T. Go, ``{The Terwilliger algebra of the hypercube},'' \href{http://dx.doi.org/https://doi.org/10.1006/eujc.2000.0514}{{\em European Journal of Combinatorics} {\bfseries 23}, 399--429 (2002)}.

\bibitem{bernard2023entanglement}
P.-A. Bernard, N.~Cramp{\'e}, and L.~Vinet, ``{Entanglement of free fermions on Hamming graphs},'' \href{http://dx.doi.org/https://doi.org/10.1016/j.nuclphysb.2022.116061}{{\em Nuclear Physics B} {\bfseries 986}, 116061 (2023)}.

\bibitem{christandl2005perfect}
M.~Christandl, N.~Datta, T.~C. Dorlas, A.~Ekert, A.~Kay, and A.~J. Landahl, ``Perfect transfer of arbitrary states in quantum spin networks,'' \href{http://dx.doi.org/https://doi.org/10.1103/PhysRevA.71.032312}{{\em Physical Review A—Atomic, Molecular, and Optical Physics} {\bfseries 71}, 032312 (2005)}.

\bibitem{farhi2024efficiently}
E.~Farhi and S.~P. Jordan, ``Efficiently constructing a quantum uniform superposition over bit strings near a binary linear code,'' \href{http://dx.doi.org/https://doi.org/10.48550/arXiv.2404.16129}{{\em arXiv preprint arXiv:2404.16129}  (2024)}.

\bibitem{li2015entanglement}
Z.-H. Li and A.-M. Wang, ``Entanglement entropy in quasi-symmetric multi-qubit states,'' \href{http://dx.doi.org/https://doi.org/10.1142/S0219749915500070}{{\em International Journal of Quantum Information} {\bfseries 13}, 1550007 (2015)}.

\bibitem{raveh2024q}
D.~Raveh and R.~I. Nepomechie, ``{$q$-analog qudit Dicke states},'' \href{http://dx.doi.org/10.1088/1751-8121/ad1ea4}{{\em Journal of Physics A: Mathematical and Theoretical} {\bfseries 57}, 065302 (2024)}.

\bibitem{koekoek2010hypergeometric}
R.~Koekoek, P.~A. Lesky, and R.~F. Swarttouw, \href{http://dx.doi.org/https://doi.org/10.1007/978-3-642-05014-5}{{\em {Hypergeometric orthogonal polynomials and their $q$-analogues}}}.
\newblock Springer, 2010.

\end{thebibliography}
\end{document}